\documentclass[aps,pra,twocolumn,amsmath,preprintnumbers,nofootinbib,secnumarabic,a4paper]{revtex4} 

\usepackage{hyperref,amsmath}\numberwithin{equation}{section}

\def\jrn#1#2#3#4#5#6{\textsc{#1}., #6, #2, \textit{#3}, \textbf{#4}, #5} \def\andd{., } 

\def\eqref#1{(\ref{#1})}
\def\scn#1#2{\section{#1}\lb{#2}}
\def\sscn#1#2{\subsection{#1}\lb{#2}}

\def\eq{Eq. } \def\eqs{Eqs. }

\def\bfl{\begin{flushleft}}
\def\efl{\end{flushleft}}
\def\bfr{\begin{flushright}}
\def\efr{\end{flushright}}
\def\bc{\begin{center}}
\def\ec{\end{center}}
\def\be{\begin{equation}}
\def\ee{\end{equation}}
\def\bse{\begin{subequations}}
\def\ese{\end{subequations}}
\def\ba{\begin{eqnarray}}
\def\ea{\end{eqnarray}}
\def\baa#1{\begin{array}{#1}}
\def\eaa{\end{array}}
\def\bw{\begin{widetext}}
\def\ew{\end{widetext}}
\def\nn{\nonumber }
\def\lb#1{\label{#1}}
\def\bit{\begin{itemize}}
\def\eit{\end{itemize}}
\def\bco{}
\def\bcs{\begin{cases}}
\def\ecs{\end{cases}}

\def\twomat#1#2#3#4{\begin{pmatrix} #1 & #2 \\ #3 & #4 \end{pmatrix}}

\def\drm{d}

\def\schrod{Schr\"odinger}

\def\pDer#1#2{\frac{\partial #1}{\partial #2}}

\def\vena{\boldsymbol{\nabla}}

\def\U{{\rm U}}

\def\cf{{\tilde D}}
\def\vol{{\cal V}}
\def\vol{V}
\def\dvol{\drm\vol}

\def\mathbb#1{#1}

\def\text#1{{\mbox{#1}}}

\def\mwf{{\cal M}}
\def\mwr{{\cal R}}
\def\mwp{{\cal S}}
\def\mwv{\boldsymbol{\cal U}}
\def\mwfo{\boldsymbol{\cal G}}
\def\mnc{\rho_0}
\def\tr{{\rm tr}}
\def\wv{\textbf{u}}
\def\wfo{\textbf{g}}
\def\nrmf{M}
\def\tb{\gamma}

\def\dn{\rho}  
\def\et{\textit{et al.}}

\begin{document}

\preprint{\small J. Theor. Appl. Mech. \textbf{57}, 843-852 (2019)   
\quad 
[\href{https://doi.org/10.15632/jtam-pl/112063}{DOI: 10.15632/jtam-pl/112063}]
}

\title{
Matrix logarithmic wave equation and multi-channel systems in fluid mechanics
}

\author{Konstantin G. Zloshchastiev}
\email{https://bit.do/kgz}
\affiliation{Institute of Systems Science, Durban University of Technology, 
P.O. Box 1334, Durban 4000, 
South Africa}


\begin{abstract} 
We formulate the mapping between a large class of nonlinear wave equations
and flow equations for barotropic fluid with internal surface tension and capillary effects.
Motivated by statistical mechanics and multi-channel physics arguments, we focus on wave equations with logarithmic nonlinearity,
and further generalize them to matrix equations.
We map the resulting equation to  flow equations of multi-channel or multi-component
Korteweg-type materials.
For some special cases,
we analytically derive Gaussian-type matrix solutions
and study them in the context of fluid mechanics.
\end{abstract}

\date{received: 14 December 2018 [JTAM], 5 June 2019 [arXiv]}


\maketitle

\section{Introduction}

Wave equations with logarithmic nonlinearity began to gain a considerable interest
among physicists since works by Rosen (1968) and Bialynicki-Birula and Mycielski (1976).
The corresponding models have been proven to be instrumental
in dealing with
physics of quantum fields
and particles
(Rosen, 1968; Rosen, 1969; Bialynicki-Birula and Mycielski, 1979; Zloshchastiev, 2010; Zloshchastiev, 2011) 
nonlinear generalizations of quantum mechanics 
(Bialynicki-Birula and Mycielski, 1976)
optics and transport or diffusion phenomena, 
nuclear physics, 
theory of dissipative systems
and quantum information 
(Yasue, 1978; Brasher, 1991; Znojil \et, 2017, Zloshchastiev, 2018a),
theory of quantum liquids and superfluidity 
(Zloshchastiev, 2011; Avdeenkov and Zloshchastiev, 2011; Zloshchastiev, 2012; Bouharia, 2015; Zloshchastiev, 2017)
and theory of
physical vacuum and classical and quantum gravity.
(Zloshchastiev, 2010; Zloshchastiev, 2011; Scott \et, 2016).

Special attention was paid to the applications
of these equations in fluid mechanics, including
the
classical hydrodynamics of Korteweg-type materials,
such as the barotropic Korteweg capillary fluids 
(De Martino \et, 2003; Lauro, 2008; Zloshchastiev, 2018b).
In such a class of models, often referred as diffuse interface models,
the capillary interface is viewed as a diffusion transition
domain of rapid smooth variation of density, while surface
tension is intrinsically incorporated 
(Dunn and Serrin, 1985; Dell'Isola F and Kosi\'nski, 1993; Anderson \et, 1998).
This allows us to describe flows with a spontaneous
nucleation, coalescence and breakdown of density inhomogeneities in
two-phase systems
(Antanovskii, 1996).
In this paper, we continue our studies in this direction,
focusing on  constructing models
which can be used to describe a variety of multi-channel
phenomena occuring in mechanics, such as multi-channel scattering, multi-component fluids and 
capillary microfluidic arrays and chips.

The paper is structured as follows.
In section \ref{s:we} we give a brief introduction to the theory of  logarithmic 
wave equations, assuming a single-channel case for simplicity.
In section \ref{s:mc} 
we consider multi-channel systems,
introduce the matrix generalization of the logarithmic 
wave equation, and formulate the basic notions of the approach.

\scn{Logarithmic wave equation: Original formulation}{s:we}
Let us consider a material (fluid being a special case), which is in thermal contact with a reservoir of infinitely large heat capacity so as to maintain 
constant temperature.
We assume that
its microscopic structure can be regarded as
a many-body system of particles, atoms and molecules, whose interaction energy is larger
than its kinetic one; 
we shall refer to such materials as condensate-like.
This class includes not only low-temperature strongly coupled systems,
but also any matter in which density or geometrical constraints make interparticle interaction potential energies
dominate over kinetic energies.

From the viewpoint of statistical mechanics,
in such materials the microstates are naturally represented 
by a canonical ensemble;
therefore
their probability $P$ is given by a standard formula,
in which kinetic energy is neglected compared to potential energy:
$ 
P \propto \exp{(-{\cal E}/T)} \approx \exp{(-\U/T)},
$ 
where $T$, ${\cal E}$ and $\U$ are the absolute temperature, energy and potential energy of a many-body system,
respectively; we shall work in the units where the Boltzmann constant $k_B = 1$.

Let us also assume  that the state of such a material can be described by a single complex-valued function
written in a Madelung form 
(Rylov, 1999):
\be\lb{e:fwf}
\Psi = \sqrt\rho \exp{(i S)}
,
\ee
where $\rho = \rho(\textbf{x},t)$ is a fluid density, and
$S = S(\textbf{x},t)$ is a phase which is related to fluid velocity: $\textbf{u} \propto \vena S$.
This function naturally obeys a normalization condition
\be\lb{e:norm}
\int_\vol |\Psi|^2 \dvol  = 
\int_\vol \rho\, \dvol = M
> 0
,\ee 
where $M$ and $\vol$ are the total mass and volume of the material.
This poses restrictions 
which are
similar to a quantum mechanical case:
the set of all normalizable functions must
constitute a Hilbert space, such as $L^2$.
Assuming that the dynamics in such space is of the Hamiltonian type,
one can derive the operator of the potential
$
\hat \U 
\sim T \ln{(A |\Psi|^2)} 
= - K (T - T_0) \ln{(\rho/\rho_0)}
$,
where $A = 1/\rho_0$ and $K$ are  
proportionality constants, 
and $\rho_0$ and $T_0$ are some critical values of, respectively, density 
and temperature at which the statistical effect vanishes
(Zloshchastiev, 2018a).

Then the energy conservation law can be recast in the form
of a logarithmic \schrod-like equation:
\ba
i \partial_t \Psi
=
\left[-\frac{\cf}{2} \vena^2
- b 
 \ln\left(|\Psi|^{2}/\rho_0\right)
\right]\Psi
,\label{e:o}
\ea
where $b = K (T - T_0)/\eta$, $\eta$ and $\cf$ are positive constants.
Here, $\eta$ has a dimensionality of a Planck constant but not necessarily its value;
from now on, we assume that $K$ is absorbed into $\eta$ for brevity.
The parameter $\cf$ can be also interpreted as
$\cf = \eta/m_\text{eff}$ where $m_\text{eff}$ is a mass of
a 
particle.

It follows also that
the nonlinear coupling must be linearly related to the thermal temperature $T$
and to the wave-mechanical temperature $T_\Psi$
\be
b \sim T \sim T_\Psi
,\ee
where the latter is defined as a thermodynamical conjugate
of the entropy function
$ 
S_\rho
=
-\int_\vol \rho \ln{\!(\rho/\rho_0)} \, \dvol
$. 
Entropy functions of this kind directly emerge from the logarithmic term
when one averages \eqref{e:o} using the inner product in a Hilbert space
of functions $\Psi$ 
(Brasher, 1991; Zloshchastiev, 2010).

\scn{Fluid-Schr\"odinger analogy}{s:sfa}
In this section, 
we derive a formal mapping between wave equations of type \eqref{e:o}
and flow equations for a certain class of materials.

\sscn{General case}{s:sfag}
Let us begin with a scalar wavefunction case (i.e., $\Psi$ is a matrix of rank one),
and consider a
general $U(1)$-symmetric Schr\"odinger equation of the form
\ba
i \partial_t \Psi
=
\left[-\frac{\cf}{2} \vena^2
- 
F (|\Psi|^{2})
\right]\Psi
,\label{e:oF}
\ea
where $F (\rho)$ is a differentiable function on a positive semi-axis $\rho$,
and $\Psi$ obeys the normalization condition \eqref{e:norm}.

By 
substituting the Madelung ansatz \eqref{e:fwf} into \eq \eqref{e:oF}, 
and separating real and imaginary parts, one obtains
\ba
\baa{l}
\partial_t\rho 
+ \cf \vena\cdot(\rho \vena S)
= 0
, 
\\
\partial_t S
- \frac{\cf}{2}
\left[
\vena\cdot \left(\frac{\vena\sqrt\rho}{\sqrt\rho}\right)
+
\frac{(\vena\sqrt\rho)^2}{\rho}
- (\vena S)^2
\right]
=F (\rho)
,
\eaa
\lb{e:flomoS}
\ea
where a dot means a scalar product. 
Taking a gradient of the latter equation
and introducing the fluid velocity 
\be
\textbf{u} = \cf \vena S
,
\ee
we obtain
hydrodynamic laws for mass and momentum conservation for a
two-phase compressible inviscid fluid with internal capillarity whose flow is
irrotational and isothermal:
\ba
\baa l 
\partial_t\rho 
+ \vena\cdot(\rho \textbf{u})
= 0
,\\ 
\partial_t \textbf{u}
+
\textbf{u} \cdot\vena \textbf{u}
-
\frac{1}{\rho} \vena\cdot \mathbb{T}
=0
,
\eaa
\lb{e:flomo}
\ea
with 
the stress tensor $\mathbb{T}$ being of the Korteweg form 
with capillary effects 
(Dunn and Serrin, 1985):
\be\lb{e:stko}
\mathbb{T} 
=
-\frac{\cf^2}{4 \rho} \vena\rho \otimes \vena\rho 
- \tilde p \, 
\mathbb{I}
,
\ee
where $\mathbb{I}$ is the identity matrix, 
and $\tilde p$
is a capillary pressure:
\be\lb{e:capp}
\tilde p 
=
p (\rho) - \frac{1}{4} \cf^2 \vena^2\rho 
,
\ee
where
\be\lb{e:eosF}
p (\rho) = 
- \cf 
\int\!\rho F'(\rho)\, \drm\rho
=
\cf
\left[
\int\! F(\rho) \drm\rho
-
\rho F(\rho)
\right]
,
\ee
is a barotropic equation of state for our fluid,
where prime means a derivative with respect to the value in brackets. 
Furthermore, the speed of sound can be derived as
\be
c_s^2
=
\pDer{\tilde p}{\dn}
=
- \cf \dn F'(\dn)
- \frac{\cf^2}{4}  
\pDer{}{\dn}\vena^2\rho
,
\lb{e:csgen}
\ee
where the second term is non-zero only if
$\vena^2\rho$ can be presented as a function of density.

Note that in the leading-order approximation with respect to $\cf$,
one can neglect the capillary term and obtain
\be\lb{e:cappapp}
\tilde p 
\approx
- \cf 
\int\!\rho F'(\rho)\, \drm\rho
,
\ \
c_s
\approx
\sqrt{
-\cf \dn
F' (\dn)
}
.
\ee

In other words, \eq \eqref{e:oF} is a concise form of writing two fluid-mechanical 
\eqs \eqref{e:flomo}.
This makes Schr\"odinger-like equations useful for studies of the Korteweg-type materials,
as well as for certain generalizations to be discussed below.

\sscn{Logarithmic fluid}{s:sfaln}
Let us apply the formulae above for a special case of a logarithmic nonlinearity.
We therefore impose
\be
F (\rho) = b \ln{\rho}
,
\ee
which corresponds to \eq \eqref{e:o}.
Then formulae 
\eqref{e:flomo}-\eqref{e:capp} remain intact,
but \eq \eqref{e:eosF} simplifies to a linear equation: 
\be
p (\rho) = - \cf b \rho
,
\ee
which belongs to a perfect-fluid class.  

Finally, in the leading-order approximation with respect to $\cf$,
one obtains
\be 
\tilde p 
\approx
- \cf b \rho
\propto (T - T_0) \dn
, \
c_s
\approx
\sqrt{
-\cf b
}
\propto
\sqrt{T - T_0}
,
\ee
which means that the logarithmic fluid is perfect,
and its speed of sound is independent of density 
(Zloshchastiev, 2011)
These two properties are useful for applying to models of those strongly-interacting fluids 
for which the perfect-fluid approximation can serve as a robust, or leading-order approximation. 


\scn{Multi-channel systems and matrix logarithmic wave equation}{s:mc}
Historically, studies of multi-channel phenomena started from multi-channel scattering in particle physics.
For instance, 
in the reaction between two nuclei such as $~^{12}$C + d, a variety of mass rearrangements
may be possible, such as $~^{13}$C + p, $~^{14}$N + $\gamma$, 
called the mass partition, 
but in all cases the
sum of the particle masses in each partition is almost the same for
each of the partitions.
Therefore, one needs to describe the whole process, taking into account not only
the existence of multiple channels but also their possible interactions (couplings) 
with each other.

In a broader context of condensed matter physics,
multi-channel phenomena can be considered through the coupled-cluster 
method (CCM) as described in the work by \v{C}\'{\i}\v{z}ek (1966).
According to this method,
the
search for a wave function is performed 
by means of the pre-multiplication ansatz,
sometimes called also a ``preconditioning'' of a wavefunction
(Acton, 1997):
 \be
 \Psi=\hat{\mwf}\,\Psi_0
,
 \label{ccma}
 \ee
where $\Psi_0$ is
a simple and often time-independent Slater determinant
reference function,
and $\hat{\mwf}$ is an operator written in an exponential form: 
$\hat{\mwf}=\exp\hat{\cal S}$. 
In quantum chemical problems, the fast convergence of the
results was reported 
(McClain \et, 2016; Hagen \et, 2016).

In essence, the coupled-cluster 
method is about reconstructing the operator  $\hat{\cal S}$
for a specific
physical state $\Psi$. 
In other words,
one
replaces constructing of a Hilbert-space state vector  by 
constructing of
a Hilbert-space operator.
The latter is usually an $N \times
N$ matrix $\hat{\mwf}$, $N \to \infty$,
but generally one also applies the truncations, \textit{i.e.}, finite-dimensional approximations $M <
\infty$. 

%

For our purposes, one can assume that 
we deal with multiple flows of the Korteweg-type with similar properties.
It implies that in a free logarithmic wave equation \eqref{e:o}
one should upgrade the function $\Psi (\textbf{x}) = \langle \textbf{x}|{\mwf} \rangle$ from a scalar function to an $N \times N$ matrix function ${\mwf}$ 
(Znojil \et, 2017):
\ba
i \partial_t \mwf
+
\frac{\cf}{2} \vena^2 \mwf
+
b 
 \ln\left( \mnc^{-1} \mwf \mwf^\dagger \right)
\mwf
= 0
,\label{e:omx}
\ea
whereas the normalization condition \eqref{e:norm} is replaced with
\be 
{\rm Tr}
({\mwf}^\dagger{\mwf})
\equiv
\langle {\mwf}|{\mwf} \rangle
=
\int_\vol {\rm tr}
({\mwf}^\dagger{\mwf})\, d V = M, 
\lb{e:normx}
\ee
where a symbol ``tr'' denotes a conventional matrix trace.
Consequently, the total mass density of such a system can be defined
as
\be 
\rho
=
{\rm tr}
({\mwf}^\dagger{\mwf}) 
,
\lb{e:denmx}
\ee
then the normalization condition \eqref{e:normx} takes a habitual form.

Furthermore,
the Madelung ansatz can be chosen in the form
\be\lb{e:fwfmx}
\mwf = 
\mwr \exp{(i \mwp)}
,
\ee
where $\mwr = \mwr (\textbf{x},t)$ and $\mwp = \mwp (\textbf{x},t)$ are self-adjoint invertible square matrices of rank $N$. 
Adjoint and invertible matrix conditions are imposed here to satisfy the correspondence principle: we must recover
a conventional nonlinear Schr\"odinger equation at $N = 1$.  
Note that $\mwr$ and $\mwp$ do not have to commute in general, and their order in \eq \eqref{e:fwfmx}
is chosen to eliminate $\mwp$ in the logarithmic term in \eqref{e:omx}.

Furthermore, matrices $\mwr$ and $\mwp$ are connected to the observables via the formulae
\ba
\rho
=
{\rm tr}
(\mwr^2) 
,
\ 
\wv
=
\cf\,
\tr
(\vena \mwp)
,
\lb{e:umx}
\ea
for which reason we call 
$\mwv = \cf \vena \mwp$ 
the velocity vector matrix, and $\mwr^2$ the mass density matrix.

Substituting the ansatz \eqref{e:fwfmx} into \eq \eqref{e:omx},
we obtain 
\ba
\baa l 
\partial_t\mwr
+ 
\vena\mwr\cdot\mwv
+
\frac{1}{2}
\mwr\vena\cdot\mwv
= 0
,\\ 
\partial_t \mwp
+ \frac{1}{2 \cf} \mwv^2
- \frac{\cf}{2} \mwr^{-1} \vena^2 \mwr
- b  \ln\left( \mnc^{-1} \mwr^2 \right)
=0
,
\eaa
\lb{e:flomomx}
\ea
where $\mwr^{-1}$ denotes the inverse matrix of $\mwr$.
From these one
can derive the following equations
\ba
\baa l 
\partial_t (\mwr^2)
+ 
\vena \cdot(\mwr^2 \mwv) 
= 0
,\\ 
\partial_t \mwv
+ (\mwv \cdot\vena)\, \mwv
- \frac{\cf^2}{2} 
\vena \left(\mwr^{-1} \vena^2 \mwr\right) \\
\qquad 
-b \cf  
\mwr^{-2} \vena\left( \mwr^2 \right)
=0
,
\eaa
\lb{e:flomomx1}
\ea
where $\mwr^{-2}$ denotes the inverse of the mass density matrix.
Here it is noticeable that the dimensional parameter $\mnc$ has disappeared 
from the equations,
indicating that the mass density and flow velocity of our material do not depend 
on a scaling constant $\mnc$.

Applying the matrix trace operation to these equations and recalling the 
definitions \eqref{e:umx},
we obtain flow equations for the observables
\ba
\baa l 
\partial_t \rho
+ 
\vena \cdot \tr (\mwr^2 \mwv) 
= 0
,\\ 
\frac{D \wv}{D t}
\equiv
\partial_t \wv
+ \tr \left[(\mwv \cdot\vena)\, \mwv \right]
=
\wfo
,
\eaa
\lb{e:flomomx2}
\ea
where 
\be\lb{e:accmx}
\wfo = \tr\, \mwfo
\ee
is a cumulative acceleration
acting on the system,
where
\be\lb{e:accmmx}
\mwfo
=
 \frac{\cf^2}{2} 
\vena \left(\mwr^{-1} \vena^2 \mwr\right)
+
b \cf\,
\mwr^{-2} \vena\left( \mwr^2 \right)
\ee
is an acceleration vector matrix.

Equations 
\eqref{e:flomomx2}  
can, of course, be recognized as  matrix analogues of the continuity and Cauchy momentum equations
for our system, respectively.
These equations can be difficult to use for actual derivation of the observables
$\rho$ and $\wv$; due to their complex structure involving products of
non-commuting matrices.
However, one should bear in mind that they are equivalent to one
matrix equation \eqref{e:omx}, which can be solved with considerably more ease.
Once its solution is found, equations 
\eqref{e:flomomx2} 
can be used for a fluid-mechanical interpretation of results.

\scn{Examples
}{s:ex}


As an example, in this section we 
consider solutions for a case when $\mwf$ is a $2\times 2$ matrix,
for reasons of simplification. 
For the same reasons, we also limit ourselves
to a single spatial dimension:  $\textbf{x} \to x$. 
Then the wave equation (\ref{e:omx}) becomes
 \be
  i \partial_t \mwf
  + 
	 \frac{1}{2} \cf\,
	\partial_{x x} \mwf +  b \ln (\mnc^{-1} \mwf \mwf^\dagger) \mwf = 0
, \label{mlnse1d}
\ee
while the normalization condition \eqref{e:normx}
becomes
\be \int\limits_{x_L}^{x_R} {\rm tr}
(\mwf^\dagger\mwf) d x = \nrmf 
, 
\lb{e:norm1d}
\ee
assuming that the system is localized in the interval of $x \in
[x_L,\,x_R]$;
this formula comes from the normalization
condition
$
   \int\limits d \vec x
 \, {\rm tr} (\mwf^\dagger\mwf)\equiv
 \langle \mwf|\mwf \rangle
 = \nrmf ,
$
where integration is taken over a spatial volume occupied by the
system, $\nrmf$ being a constant usually interpreted as a number of
particles inside such a volume.

Furthermore, below we consider the cases when the exact
analytical solutions of \eq  (\ref{mlnse1d}) are known.

\subsection{Diagonal case}

Let us assume the simplest case when $\mwf$ is imposed to be a diagonal $2\times 2$
matrix.
The imposed diagonality indicates that the cross-channel 
coupling terms can be neglected,
\textit{i.e.}, our configuration or material is a mixture of two independent components. 
The ground-state
solution of \eq (\ref{mlnse1d}) can be found analytically.
After some algebra, we obtain
 \ba
\mwf
&= &
\sqrt{\mnc} \twomat{\psi_{1}({x},t)}{0}{0}{\psi_{2}({x},t)} 
\nn\\
&=&
\twomat{R_{1}({x},t)}{0}{0}{R_{2}({x},t)}\,
\exp\!\twomat{- i E_1 t}{0}{0}{- i E_2 t}\!,~~~~~
\label{e:psidiag}
 \ea
where $\psi_{a}({x},t)$ are complex-valued functions.  
In the fluid rest
frame of reference, the solution has the form of the gausson, \textit{i.e.}, the Gaussian parcel
modulated by de Broglie plane wave,
 \be\label{e:psidiagf} \psi_a
(x,t) = C_a \exp{\left(- \frac{1}{2} \tb\, x^2  + \nu_a x - i E_a
t\right)} ,
 \ee
where $a=1,\,2$, $\tb = 2 b/\cf > 0$, and 
 \be
 E_a = b (1 - \ln{C_a^2}) - \frac{1}{2} \cf \, \nu_a^2
  \ee
is energy of a wave in the $a$th channel. 
The  integration constants $C_a$ and
$\nu_a$ are related, together with $b$, to a
mean and variance of the solution.
Note that this solution exists for models with a positive value of $\tb$,
as discussed in the concluding section.

If one imposes also a normalization condition (\ref{e:norm1d}) here,
then one obtains an additional constraint for the integration constants
\be\label{e:normco}
\Delta F_1 + \Delta F_2 = - 2 \sqrt{\frac{\tb}{\pi}} \frac{\nrmf}{\mnc}
,
\ee
where we denoted
$\Delta F_a = F_a (x_R) - F_a (x_L)$
and
$F_a (x) = C_a^2\, \exp{\left(\frac{\nu_a^2}{\tb}\right)}\, \text{erf}\left(\frac{\nu_a - \tb x}{\sqrt{\tb}}\right)$.
For instance,
when $x_R$ and $x_L$ are set to, respectively, plus and minus infinity,
then this constraint takes a simple form
$
\sum\limits_{a=1}^2
C_a^2\, \exp{\left(\frac{\nu_a^2}{\tb}\right)}
= \sqrt{\frac{\tb}{\pi}} \frac{\nrmf}{\mnc}
$.

Note that due to the Galilean symmetry of a logarithmic \schrod~equation, 
one can always generate solitary wave solutions from a solution
(\ref{e:psidiagf}), whose
centers of mass propagate with velocity $v_a$, independently for
each channel. 
One can check also that the solution (\ref{e:psidiag})
naturally has a commutative property,
$ 
 \left[\mwf^\dagger,\, \mwf \right] = 0 $,
thus eliminating the ordering uncertainty
inside a logarithmic term. 

Finally, one can verify that the cumulative acceleration \eqref{e:accmx}
and acceleration matrix \eqref{e:accmmx} both vanish on the solution \eqref{e:psidiag}, \eqref{e:psidiagf}:
\be
(\mwfo)_x
= 
\twomat{0}{0}{0}{0}
, \ \
\wfo = \textbf{0}
,
\ee
which means that the material derivative of flow velocity is zero for this solution.

\subsection{Off-diagonal case}

Now let us assume that the
channel-coupling terms are dominating. 
Therefore, one can
impose that $\mwf$ is an off-diagonal matrix 
 \ba
\mwf
&=& 
\sqrt{\mnc} 
\twomat{0}{\bar\psi_{2}({x},t)}{\bar\psi_{1}({x},t)}{0}
\nn\\
&=&
\twomat{0}{\bar R_{2}({x},t)}{\bar R_{1}({x},t)}{0}\,
\exp\!\twomat{- i \bar E_1 t}{0}{0}{- i \bar E_2 t}\!,
~~~~~~
\label{e:psiodiag}
 \ea
where $\bar\psi_{a}({x},t)$ are
complex-valued functions.

The ground-state solution of \eq (\ref{mlnse1d}) can be found
analytically. 
In a rest frame, it also has the form of the Gaussian
 modulated by de Broglie plane wave,
 \be\label{e:psiodiagf} \bar\psi_a
(x,t) = C_a \exp{\left(- \frac{1}{2} \tb\, x^2  + \nu_a x - i \bar E_a
t\right)} ,
 \ee
where $a=1,\,2$, 
and 
 \be
 \bar E_a = b (1 - \ln{C_a^2}) - \frac{1}{2} \cf \, \nu_a^2
  \ee
is energy of a wave for the $a$th channel, $C_a$ and $\nu_a$ are integration constants.
As in the diagonal case above, this solution exists for models with a positive value of $\tb$,
as discussed in the concluding section. 

If one imposes also a normalization condition
(\ref{e:norm1d}) then one obtains an additional constraint for the
integration constants, which is identical to \eq (\ref{e:normco}).


Finally, 
acceleration matrix \eqref{e:accmmx} 
takes a non-zero value
on the solution \eqref{e:psiodiag}, \eqref{e:psiodiagf},
given by:
\be 
(\mwfo)_x
= b \cf (\nu_2 - \nu_1)
\twomat{1}{0}{0}{-1}
, 
\label{e:accmmxodiag} \ee 
but
the cumulative acceleration \eqref{e:accmx}
still vanishes, thus indicating
that the material derivative of flow velocity is zero for this solution.

\scn{Conclusion}{s:con}

In this paper, we formulated a formal mapping between a large class of nonlinear wave equations
and flow equations of a barotropic fluid with internal surface tension and capillary effects.
Then we
generalized quantum wave equations of
a logarithmic type to a matrix form.
Furthermore, we mapped this equation to matrix flow equations of  multi-channel or multi-component
Korteweg-type materials including fluids with internal surface tension and capillary effects.

Aside from being a concise way of writing flow equations,
such mapping
reveals the existence of a Hilbert space of states represented by
different solutions, for the same set of boundary
conditions.
This means that the material can randomly ``choose'' either of these states, so that
a statistical ensemble occurs.

Furthermore,
in Section \ref{s:ex}, 
we studied two simplest examples of the model, namely,
a case of $2\times 2$ matrices in one-dimensional space,
assuming a positive value of $\tb$ and either a diagonal or off-diagonal form.
We analytically derived Gaussian-type matrix solutions,
and studied them in the context of fluid mechanics.
It should be emphasized that the solutions presented 
above exist only for models with
positive values of $\tb$ where they usually represent states
with lowest energy - an analogue of ground states in quantum mechanics.
For negative values of $\tb$, the topology of the problem
changes crucially, and the topological solitons emerge,
as was shown for the scalar (rank-1 matrix) case 
(Zloshchastiev, 2018b).

We expect that these matrix fluid models will find application not only in the
theory of low-temperature superfluids, but
also for mechanical studies of arrays of microfluids and other multi-channel flows where capillary effects are substantial.


\begin{acknowledgments}
I am grateful to participants of the XXIII Fluid Mechanics Conference KKMP-2018 in
Zawiercie, Poland (9-12 September 2018),
where preliminaries of this work were reported and discussed. 
Proofreading of the manuscript by P. Stannard is greatly appreciated as well.
This work is based on research supported by the Department of Higher Education and Training of South Africa,
as well as by
the National Research Foundation of South Africa under Grant No. 95965.

\end{acknowledgments}



\def\PR{Physical Review}
\def\JMP{Journal of Mathematical Physics}
\def\CMP{Communications in Mathematical Physics}
\def\APNY{Annals of Physics}
\def\PL{Physics Letters}
\def\PSc{Physica Scripta}
\def\IJTP{International Journal of Theoretical Physics}
\def\GC{Gravitation \& Cosmology}
\def\APP{Acta Physica Polonica}
\def\ZN{Zeitschrift f\"ur Naturforschung}
\def\JPB{Journal of Physics B: Atomic, Molecular and Optical Physics}
\def\EPJ{European Physical Journal}
\def\EPL{Europhysics Letters (EPL)}
\def\GAFD{Geophysical \& Astrophysical Fluid Dynamics}
\def\ARMA{Archive for Rational Mechanics and Analysis}
\def\ARFM{Annual Review of Fluid Mechanics}
\def\JPCS{Journal of Physics: Conference Series}

\end{document}